# Comment on "High-Pressure Behavior of Hydrogen and Deuterium at Low Temperatures"

Alexander F. Goncharov [1] and Yu. A. Freiman[2]

[1]Geophysical Laboratory, Carnegie Institution of Washington, 5251 Broad Branch Road NW, Washington D.C. 20015, USA
[2]B. Verkin Institute for Low Temperature Physics and Engineering of the National Academy of Sciences of Ukraine, 47 Lenin Avenue, Kharkov, 61103, Ukraine

**Abstract**
X.-D. Liu et al. [Phys. Rev. Lett. 119, 065301 (2017)] report on the existence of a new unique solid phase of $D_2$, which makes the high-pressure low-temperature behavior distinct from $H_2$. Here, based on the analysis of their Raman data and phase transition theory, we show that the presented data do not support this claim.

**Main text**
High-pressure behavior of hydrogen is of great interest because of theoretically predicted metallization and formation of exotic dissipation-free phases [1]. While fully refined crystal structure of hydrogen is only known for phase I, where freely rotating $H_2$ molecules form an *hcp* lattice [2], Raman spectroscopy can in principle probe phase changes due to sensitivity of the vibrational selection rules to the molecular orientational order. For the vibron mode, this may result in a discontinuous change in frequency at the transition and temperature dependent vibron frequency in a low-symmetry phase II where the orientation ordering (described via the order parameter [3, 4]) is progressing on cooling down (or pressure increase). However, no substantial temperature dependence of the vibron mode is expected in phase I, which is rotationally disordered.

In contrast to these simple expectations, Liu et al. [5] report surprising minima in the temperature dependencies of vibron modes in both $H_2$ and $D_2$ (cf. Refs. [6-8]). In $D_2$ they observe these minima above 112 GPa, while at lower pressures a step-like positive increase of the vibron frequency is reported in fair agreement with other experiments [8, 9]. They also present an additional roton peak and a change in slope in the pressure dependence of the Raman vibron sidebands to support their claim of the existence of a distinct broken symmetry phase II'. These latter observations are technically questionable. Indeed, the additional roton peak could appear as due to lifting of the accidental degeneracy under pressure while a change in slope is not obvious from the data presented and has not been statistically proven in Ref. [5].

The vibron dips are small, the magnitude is just 1-2 cm$^{-1}$ for $D_2$, which is comparable to the spectral resolution of the experiments and the accuracy of the peak position determination. It is unclear whether the measured vibron mode is a single peak or a doublet (*e.g.* due to phase coexistence [6]) as the vibron discontinuity can be smaller than the spectral resolution. One cannot judge about this as the spectra are not presented and there is no peak position/linewidth analysis in Ref. [5]. We propose that there is a positive abrupt change in the vibron frequency at the II-I transition (Fig. 1). In the case the Raman vibron has a strong temperature dependence in phase II, this will result in the dips similar to those reported in Ref. [5]. In the case the vibron frequency of the broken symmetry phase is weakly temperature dependent (<112 GPa), the dips cannot be observed.

Our analysis shows that the distinct phases II and II' of Ref. [5] differ by the temperature dependence of the Raman vibron in the broken symmetry phase, which is not directly related to the claimed change of symmetry. In this regard, please also note that the data of Ref. [5] are not sufficiently dense to claim a puzzling dip in the phase line near the proposed triple point between phases II, II', and I at 112 GPa and 110 K. Thus, we believe that the available data are insufficient to claim the existence of a new broken symmetry phase II'. Moreover, the proposed nomenclature is confusing because another phase II' was reported [6], the existence of which the authors of Ref. [5] could not refute nor support because their measurements lack sufficient spectral and P-T resolution.

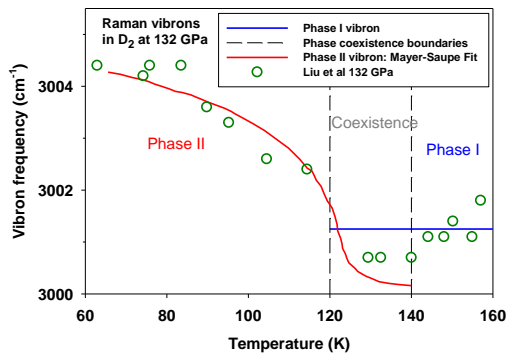

Fig. 1. Observations of the dips in the temperature dependent vibron Raman spectra of $D_2$ at 132 GPa [5]. Red curve is a manual fit to the data using the Mayer-Saupe order parameter [3].